\PassOptionsToPackage{table,xcdraw}{xcolor}
\documentclass[sigconf]{acmart}
\usepackage{textcomp}
\let\oldftcp\footnotetextcopyrightpermission
\renewcommand\footnotetextcopyrightpermission[1]{\oldftcp{%
}}

\AtBeginDocument{%
  }

\acmConference[ACM REP’23]{The ACM Conference on Reproducibility and Replicability}{June 27--29,2023}{Santa Cruz, California}

\usepackage{graphicx}

\usepackage{multirow}
\usepackage{caption}
\usepackage{subcaption}
\usepackage{lipsum} 
\usepackage{listings}
\usepackage{framed}
\usepackage{amsmath}
\usepackage{balance}

\usepackage{enumerate}
\usepackage[shortlabels]{enumitem}

\definecolor{codepurple}{rgb}{0.58,0,0.82}
\definecolor{backcolour}{RGB}{239, 239, 239}
\definecolor{codeorange}{RGB}{191, 94, 45}
\definecolor{codeblue}{RGB}{0, 0, 255}
\definecolor{codegreenl}{rgb}{0,0.6,0}
\definecolor{codegreend}{RGB}{101, 139, 111}
\definecolor{codegray}{rgb}{0.5,0.5,0.5}

\lstdefinestyle{mystyle}{
    backgroundcolor=\color{backcolour},   
    commentstyle=\color{codepurple},
    keywordstyle=\color{codeorange},
    numberstyle=\small\color{codegray},
    stringstyle=\color{codegreenl},
    basicstyle=\fontsize{8}{8}\selectfont\ttfamily,
    breakatwhitespace=false,     
    breaklines=true,                 
    captionpos=b,                    
    keepspaces=true,                 
    numbers=none,                    
    numbersep=2pt,        
    numbers=left,
    stepnumber=1,
    showspaces=false,                
    showstringspaces=false,
    showtabs=false,                  
    tabsize=2
}
\begin{document}


\title{KheOps: Cost-effective Repeatability, Reproducibility, and Replicability of Edge-to-Cloud Experiments}


\author{Daniel Rosendo}
\email{daniel.rosendo@inria.fr}
\affiliation{
 \institution{Univ Rennes, Inria, CNRS, , IRISA}
 \city{Rennes}
 \country{France}}

\author{Kate Keahey}
\email{keahey@mcs.anl.gov}
\affiliation{
 \institution{Argonne National Laboratory}
 \city{Chicago}
 \country{USA}}

\author{Alexandru Costan}
\email{alexandru.costan@inria.fr}
\affiliation{
 \institution{Univ Rennes, Inria, CNRS, IRISA}
 \city{Rennes}
 \country{France}}

\author{Matthieu Simonin}
\email{matthieu.simonin@inria.fr}
\affiliation{
\institution{Univ Rennes, Inria, CNRS, IRISA}
 \city{Rennes}
 \country{France}}

\author{Patrick Valduriez}
\email{patrick.valduriez@inria.fr}
\affiliation{
 \institution{Univ Montpellier, Inria, CNRS, LIRMM}
 \city{Montpellier}
 \country{France}}

\author{Gabriel Antoniu}
\email{gabriel.antoniu@inria.fr}
\affiliation{
 \institution{Univ Rennes, Inria, CNRS, IRISA}
 \city{Rennes}
 \country{France}}


\begin{abstract}
Distributed infrastructures for computation and analytics are now evolving towards an interconnected ecosystem allowing complex scientific workflows to be executed across hybrid systems spanning from IoT Edge devices to Clouds, and sometimes to supercomputers (the Computing Continuum). Understanding the performance trade-offs of large-scale workflows deployed on such complex Edge-to-Cloud Continuum is challenging. To achieve this, one needs to systematically perform experiments, to enable their reproducibility and allow other researchers to replicate the study and the obtained conclusions on different infrastructures. This breaks down to the tedious process of reconciling the numerous experimental requirements and constraints with low-level infrastructure design choices. 

To address the limitations of the main state-of-the-art approaches for distributed, collaborative experimentation, such as Google Colab, Kaggle, and Code Ocean, we propose KheOps, a collaborative environment specifically designed to enable cost-effective reproducibility and replicability of Edge-to-Cloud experiments. KheOps is composed of three core elements: (1) an experiment repository; (2) a notebook environment; and (3) a multi-platform experiment methodology.


We illustrate KheOps with a real-life Edge-to-Cloud application. The evaluations explore the point of view of the authors of an experiment described in an article (who aim to make their experiments reproducible) and the perspective of their readers (who aim to replicate the experiment). The results show how KheOps helps authors to systematically perform repeatable and reproducible experiments on the Grid5000 + FIT IoT LAB testbeds. Furthermore, KheOps helps readers to cost-effectively replicate authors experiments in different infrastructures such as Chameleon Cloud + CHI@Edge testbeds, and obtain the same conclusions with high accuracies ($>$88\% for all performance metrics).
\end{abstract}

\ccsdesc[500]{Computing methodologies}
\ccsdesc[500]{Computing methodologies~Distributed computing methodologies}
\ccsdesc[500]{General and reference~Experimentation}
\ccsdesc[500]{General and reference~Measurement}

\keywords{Reproducibility, Replicability, Repeatability, Computing Continuum, Workflows, Edge Computing, Cloud Computing}


\maketitle

\section{Introduction}
\label{sec:introduction}

Modern scientific workflows require hybrid infrastructures, combining resources and services executed on the IoT/Edge with other resources and services running on Clouds or on HPC systems (the \emph{Computing Continuum}~\cite{etp4-hpc-20}) to enable their optimized execution. 
Due to the complexity of application deployments on such highly distributed and heterogeneous Edge-to-Cloud infrastructures, realizing the Computing Continuum vision in practice remains burdensome. 

\begin{figure}[t]
    \centering
         \includegraphics[width=0.85\linewidth]{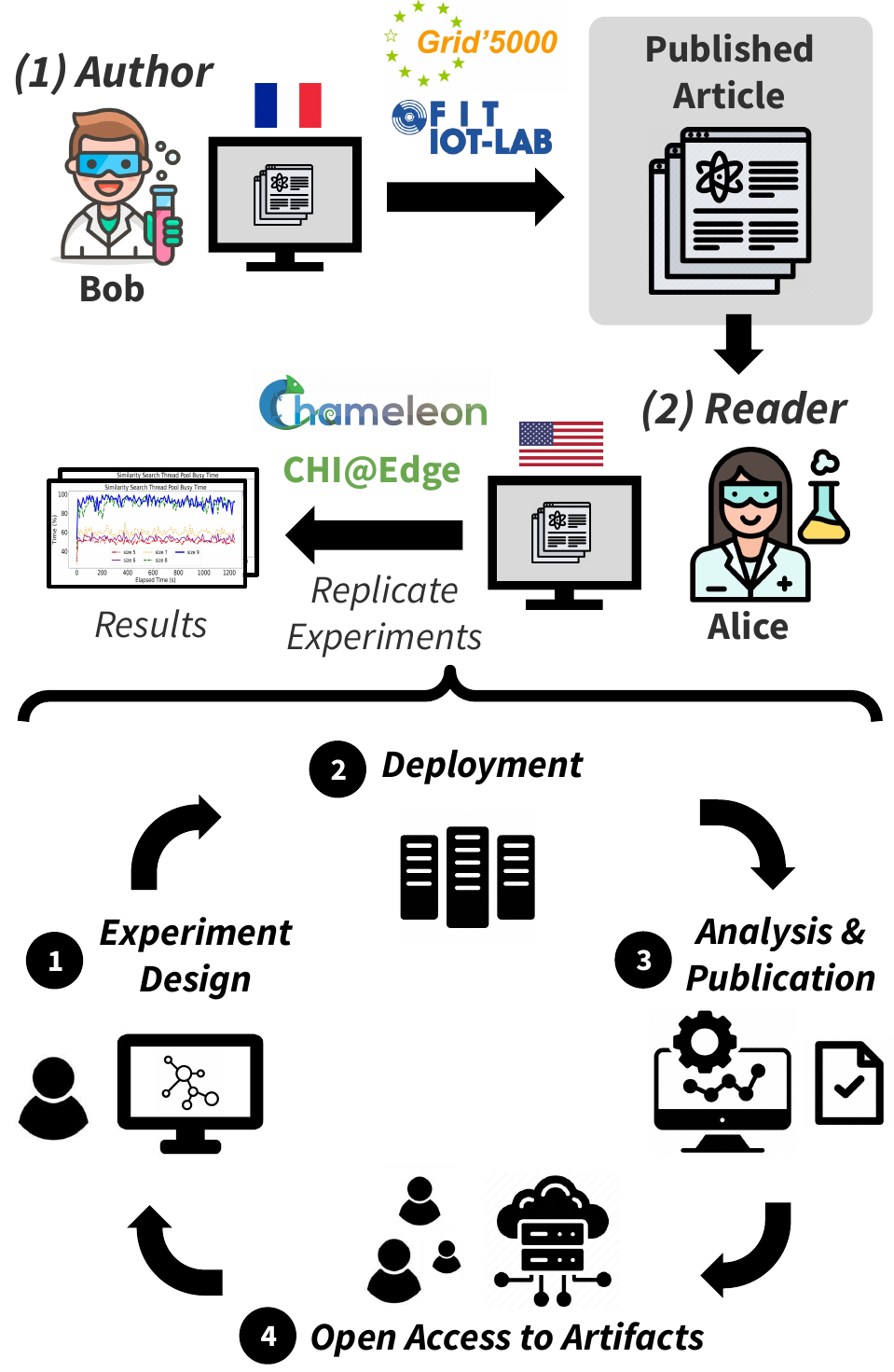}
    \caption{Processes for reproducing and replicating experiments regarding the authors and readers point of view.}
    \label{fig:context}
\end{figure}

One challenge stems from systematically performing experiments on the continuum. In particular, the processes enabling their reproducibility, as well as the replication of the performance trade-offs are inherently difficult~\cite{haibe2020transparency}. Figure~\ref{fig:context} illustrates such processes. Let us consider the case of a group of researchers who execute their experiments on French scientific testbeds such as Grid'5000~\cite{RaphaEtAl2006} (providing Cloud/HPC servers) and FIT IoT LAB~\cite{adjih2015fit} (providing IoT/Edge devices), and want to publish their results in an article. Next, the readers want to replicate the experiments on American testbeds such as the Chameleon Cloud~\cite{KateEtAl2020} and CHI@Edge~\cite{keahey2021chameleon}.

These processes compel a lot of effort, are time-consuming, and bring many technical challenges for both sides. For instance, also depicted in Figure~\ref{fig:context}, they require: (1) following methodologies to systematically design the experiments and to reconcile many application requirements or constraints in terms of energy consumption, network efficiency, and hardware resource usage; (2) configuring systems and networks, and deploying applications on testbeds for large-scale evaluations; (3) analyzing, repeating experiments, and publishing results; and (4) finally, providing open access to the experiment artifacts in a public and safe repository.

\begin{figure}[t]
  \centering
  \includegraphics[width=0.95\linewidth]{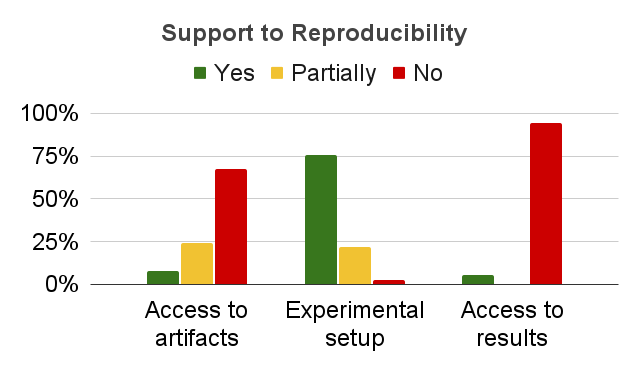}
  \caption{Support to the reproducibility of Edge-to-Cloud experiments provided by the 34 studies in our survey~\cite{rosendo:hal-03654722}.}
  \label{fig:exp-repro}
\end{figure}

Given such complexities, researchers end up not following rigorous methodologies for supporting the reproducibility of the experiments, as observed in our previous survey~\cite{rosendo:hal-03654722} and summarized in Figure~\ref{fig:exp-repro}. As a consequence, it makes it hard for other researchers to replicate the published studies~\cite{krafczyk2021learning}.

Let us sum up the associated requirements in this context~\cite{stodden2016enhancing, gundersen2018reproducible}. To enable \textbf{reproducible} experiments on the Edge-to-Cloud continuum, the requirements (a-REQ) of the authors of the experiments can be described as follows:

\begin{enumerate}[start=1,label={\textbf{a-REQ \arabic*.}}]
    \item Execute experiments on heterogeneous computing resources (\emph{e.g.,} IoT/Edge and Cloud/HPC infrastructures).   
    \item Systematically describe and explain the experimental processes and their reasoning.
    \item Efficiently configure the experimental infrastructure and express topologies in repeatable ways.
    \item Easily share the experiment artifacts in a public and safe repository.
\end{enumerate}

At the same time, to enable the \textbf{replicability} of the experiments, the readers of an article describing those experiments have the following requirements (r-REQ):

\begin{enumerate}[start=1,label={\textbf{r-REQ \arabic*.}}]
    \item Find and access the experiment as simply as finding and reading its paper.
    \item Perform the experiment, not just read about it.
    \item Answer not just to the \emph{“What”} question (What the experiment does?), but also the \emph{“Why”} (Why did authors set up that way?) and \emph{“How”} (How did authors connect machines/devices?)
    \item Efficiently configure the experimental infrastructure to reduce the time spent satisfying all the experiment requirements.
\end{enumerate}

In this paper, we study the challenges of reproducing and replicating Edge-to-Cloud experiments in cost-effective ways. \textbf{Cost-effective means} to allow authors and readers to easily fulfill their experimental requirements as previously described. This calls for practical solutions beyond the state-of-the-art. 

Our main objective is to provide a collaborative environment and methodology that supports reproducible Edge-to-Coud experimentation between different open testbeds such as Grid'5000, FIT IoT LAB, Chameleon, \emph{etc.}, equipped to deal with IoT/Edge and Cloud/HPC resources which are fundamental to reproducibility~\cite{keahey2020silver}. We propose the following main contributions:

\begin{enumerate}

    \item \textbf{A study of the characteristics of the main state-of-the-art collaborative environments} (\emph{e.g.,} Google Colab, Kaggle, and Code Ocean) for enabling reproducible experiments. Their \textbf{main limitations in the context of Computing Continuum research} are discussed in Section~\ref{sec:limitations}.

    \item \textbf{A novel collaborative environment to enable reproducible Edge-to-Cloud experiments} (Section~\ref{sec:kheops}). This approach, named KheOps, allows researchers to reproduce and replicate Edge-to-Cloud workflows cost-effectively. KheOps core elements are: (1) a portal for sharing experiment artifacts; (2) a notebook environment for packaging code, data, environment, and results; and (3) a \textbf{multi-platform} experimental methodology for deploying experiments on heterogeneous resources from the IoT/Edge (FIT IoT LAB and CHI@Edge) to the Cloud/HPC Continuum (Grid5000 and Chameleon). We highlight that KheOps may be integrated with other large-scale scientific testbeds. 

    \item An \textbf{experimental validation} of the proposed approach with a \textbf{real-world use case deployed on real-life IoT/Edge devices and Cloud/HPC systems}. The evaluations show that KheOps helps: (1) authors to perform reproducible experiments on the \textbf{Grid5000 + FIT IoT LAB} testbeds, and (2) readers to cost-effectively replicate authors experiments on the \textbf{Chameleon Cloud + CHI@Edge} testbeds, and \textbf{obtain the same conclusions with high accuracies}, $>$88\% for all performance metrics (Section~\ref{sec:evaluation}).
    
\end{enumerate}

\section{Background}
\label{sec:background}

\begin{table}[t]
\fontsize{8.5}{12}\selectfont
\centering
\caption{ACM Digital Library Terminology Version 1.1 \cite{ferro2018sigir}}
\label{tbl:acm-terminology}
\begin{tabular}{ll}
\hline
\rowcolor[HTML]{e2eaec}
\parbox[t]{2mm}{\rotatebox[origin=c]{90}{\textbf{\hspace{0.3cm}Repeatability\hspace{0.3cm}}}} & 
\textit{\begin{tabular}[c]{@{}l@{}} Same team, same experimental setup: the measurement can \\be obtained with stated precision by the same team using the \\same measurement procedure, the same measuring system, \\under the same operating conditions, in the same location on \\multiple trials. For computational experiments, this means \\that a researcher can reliably repeat their own computation.\end{tabular}}
\\
\hline
\parbox[t]{2mm}{\rotatebox[origin=c]{90}
{\textbf{\hspace{0.3cm}Reproducibility\hspace{0.3cm}}}} &
\textit{\begin{tabular}[c]{@{}l@{}} Different team, same experimental setup:  the measurement \\can be obtained with stated precision by a different team using \\the same measurement procedure, the same measuring system, \\under the same operating conditions, in the same or a different \\location on multiple trials. For computational experiments, this \\means that an independent group can obtain the same result \\using the author’s own artifacts.\end{tabular}}
\\ 
\hline
\rowcolor[HTML]{e2eaec}
\parbox[t]{2mm}{\rotatebox[origin=c]{90}
{\textbf{\hspace{0.6cm}Replicability\hspace{0.6cm}}}} &
\textit{\begin{tabular}[c]{@{}l@{}} Different team, different experimental setup: the measurement \\can be obtained with stated precision by a different team, a \\different measuring system, in a different location on multiple \\trials. For computational experiments, this means that an \\independent group can obtain the same result using artifacts \\which they develop completely independently.\end{tabular}}
\\
\hline
\end{tabular}
\end{table}

In this section, we start by defining the terms repeatability, reproducibility, and replicability (Section~\ref{subsec:3rs-def}). Next, we explore the following research question from the  Computing Continuum perspective: \emph{\textbf{What would a good collaborative system look like?}} In our vision, it should: (1) allow users to share research artifacts in a public and safe repository (Section~\ref{subsec:artifacts}); (2) provide an environment for setting up and describing experiments step-by-step (Section~\ref{subsec:environment}); (3) provide experimental methodologies to leverage  heterogeneous Edge-to-Cloud computing resources from various scientific testbeds, at large-scale (Section~\ref{subsec:methodology}).

\subsection{Repeatability, Reproducibility, Replicability}
\label{subsec:3rs-def}

An important requirement for researchers from various communities is that the scientific claims be verifiable by others (\emph{i.e.,} building upon published results). As illustrated in Figure~\ref{fig:exp-repro}, such requirement is hardly satisfied in the context of Computing Continuum experiments. This can be achieved through repeatability, reproducibility, and replicability (3Rs)~\cite{ieee-computing, Stodden-2014}.  There are many non-uniform definitions of the 3Rs in literature. In this work, we follow the terminology proposed by the ACM Digital Library \cite{ferro2018sigir} (Artifact Review and Badging version 1.1), as presented in Table \ref{tbl:acm-terminology}. 

Achieving \textbf{repeatability} means that one can reliably repeat the experiments and obtain precise measurements (\emph{e.g.,} Edge to Cloud processing latency, memory consumption, among others) by using the same methodology and artifacts (\emph{i.e.,} same testbed, same physical machines, same libraries/framework, same network configuration). Executing multiple experiments allows us to explore different scenario settings (\emph{e.g.,} varying the number of Edge devices) and explore the impact of various parameters (\emph{e.g.,} the network configuration between Edge devices and the Cloud server) on the performance metrics. 

\textbf{Reproducibility} means that external researchers having access to the original methodology (\emph{e.g.,} configuration of physical machines, network and systems, scenario descriptions) and using their own artifacts (\emph{i.e.,} data sets, scripts, AI frameworks, \emph{etc.}) can obtain precise measurements of the application processing latency and throughput, for instance.

\textbf{Replicability} refers to independent researchers (\emph{i.e.,} the readers of an article that was published by a different team) having access to the original methodology and artifacts (\emph{e.g.,} configuration of physical machines, processing steps, network setup, \emph{etc.}) and performing the experiments in different testbeds. The goal is that independent researchers can obtain precise results and conclusions consistent with the original study.

\subsection{Trovi sharing portal}
\label{subsec:artifacts}

Collaborative systems should be integrated with public and safe repositories providing open access to the research artifacts to enable the reproducibility of experiments. Repositories like Trovi~\cite{trovi}, Kaggle~\cite{kaggle}, Code Ocean Explorer~\cite{code-ocean}, AI Hub~\cite{aihub}, GitHub~\cite{github}, and Zenodo~\cite{zenodo} allow users to store versioned and citeable (\emph{e.g.,} through a DOI: Digital Object Identifier) artifacts such as code, datasets, or Jupyter notebooks, among others. 

In this work, we leverage on the Trovi sharing portal because it provides a public REST API that facilitates integration with existing systems. Furthermore, Trovi provides a series of features to manage research artifacts such as: integration with GitHub and Zenodo; creating, packaging, and sharing artifacts as Jupyter notebooks with 500MB in total size by default; support for scientific testbeds like Chameleon, which allows users to re-launch the available artifacts on the testbed.

\subsection{Jupyter environment}
\label{subsec:environment}

Another important aspect for reproducible and replicable experiments is that collaborative systems support executable research packages composed of code, data, environment configurations, and experiment results. The most popular open-source solutions are Jupyter notebooks~\cite{kluyver2016jupyter} and Apache Zeppelin~\cite{zeppelin}. In this work, we use Jupyter notebooks for packaging research artifacts due to its wider compatibility with operating systems and programming languages, and the community support.

The Jupyter project consists of JupyterHub, JupyterLab, and notebooks. JupyterHub aims to serve Cloud-based Jupyter notebooks for multiple users. The goal is to provide users a ready-to-use computational environment with their own workspace on shared resources. JupyterHub servers are customizable, scalable, and portable on a variety of infrastructures. It is composed of a Hub that manages the following sub-services: a proxy that receives requests from clients; spawners to monitor notebook servers; and an authenticator to manage how users access the system.

JupyterLab refers to a web-based user interface providing mainly: notebook, terminal, text editor, file browser, and rich outputs. It allows users to configure and arrange their experimental workflows, as well as adding extensions to expand and enrich functionalities.

Finally, notebooks allow users to create programming documents combining: (1) formatted text (\emph{e.g.,} \textit{prospective
data} that explains each step of an experiment workflow); (2) executable code with the respective outputs (\emph{e.g., \textit{retrospective data} derived by the execution}); and (3) experimental results with visualizations and various sorts of rich media, such as images and videos.

\subsection{E2Clab experimental methodology}
\label{subsec:methodology}

Understanding and optimizing workflow performance requires executing and reproducing complex experiments at large scale. Several existing environments aid users to run such experiments. Their limitations are discussed in the next section and summarized in Table~\ref{tbl:limitations}. Based on these findings and the specific Computing Continuum requirements, in this work we leverage the E2Clab methodology.

\begin{figure}[t]
    \centering
         \includegraphics[width=\linewidth]{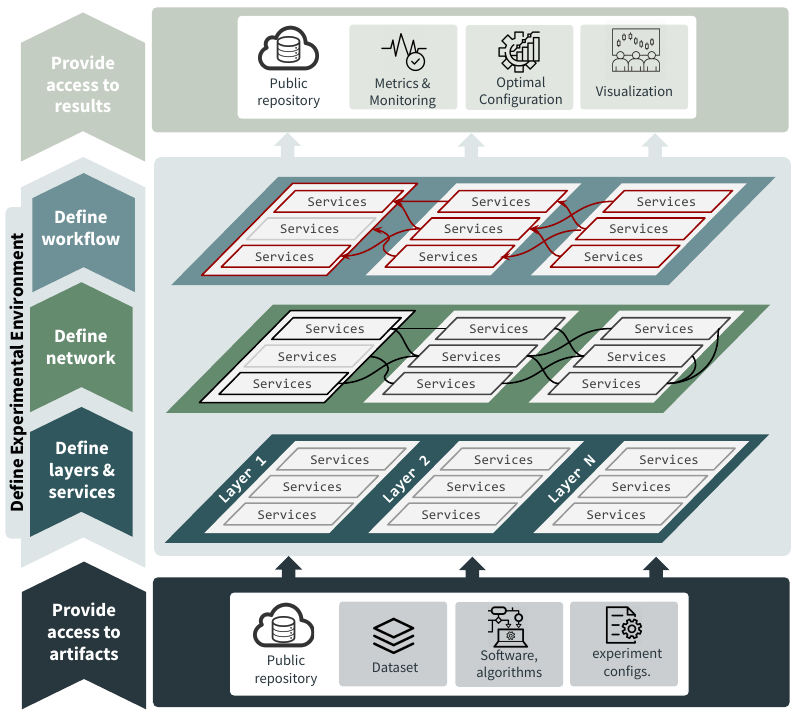}
    \caption{\textbf{E2C}\textit{lab} experiment methodology~\cite{rosendo:hal-02916032}.}
    \label{fig:methodology}
\end{figure}

E2Clab~\cite{rosendo:hal-02916032} is an open-source framework (available at~\cite{e2clab-code}) that implements a rigorous methodology (illustrated in Figure~\ref{fig:methodology}) for designing experiments with real-world workloads on the Edge-to-Cloud Continuum. It allows researchers to reproduce the application behavior in a controlled environment in order to understand and optimize performance~\cite{rosendo:hal-03310540}. E2Clab sits on top of EnOSlib~\cite{cherrueau2021enoslib} to enforce the experiment configurations on testbeds. High-level features provided by E2Clab are: \emph{(i)} reproducible experiments; \emph{(ii)} mapping application parts (Edge, Fog and Cloud/HPC) and physical testbeds; \emph{(iii)} experiment variation and transparent scaling of scenarios; \emph{(iv)} defining Edge-to-Cloud network constraints; \emph{(v)} experiment deployment, execution and monitoring (\emph{e.g.,} on Grid'5000, Chameleon, and FIT IoT LAB); and \emph{(vi)} workflow optimization.

\section{Limitations of Existing Collaborative Environments}
\label{sec:limitations}

\begin{table*}[t]
\small
\centering
\caption{Limitations of Existing Collaborative Environments.}
\label{tbl:limitations}
\begin{tabular}{m{2.5cm}m{4.6cm}m{4.6cm}m{4.6cm}}
\hline
\textbf{\begin{tabular}[c]{@{}r@{}} \end{tabular}}
\textbf{Limitation} & \textbf{Google Colab} & \textbf{Code Ocean} & \textbf{Kaggle}                                         \\ \hline

\rowcolor[HTML]{D9D9D9} 
\textbf{\begin{tabular}[c]{@{}l@{}}Resource\\ heterogeneity \end{tabular}} 
& 
CPU, disk, and memory limits; GPU types available; no access to IoT/Edge devices; &
experiments run on AWS virtual machines; no access to IoT/Edge devices; &
limits CPU, GPU, and TPU access; does not support IoT/Edge devices;
\\

\textbf{\begin{tabular}[c]{@{}l@{}}Large-scale\\ experiments\end{tabular}} & 
limits sessions to 12 hours; paid access to multiple computing resources. &
limits access to 10 compute hours; paid access to multiple computing resources. &
limits execution time to 12 hours; paid access to Google Cloud Services.
\\
\rowcolor[HTML]{D9D9D9} 
\textbf{\begin{tabular}[c]{@{}l@{}}Repeatability, \\Reproducibility, \\Replicability \end{tabular}} & 
hard to repeat and reproduce experiments on the same hardware: resource availability varies over time and usage limits fluctuate. Replicability in different infrastructures (\emph{e.g.,} beyond Google machines) is not straightforward. 
&
lacks support for the reproducibility of distributed experiments. Computing and storage resources are available in AWS virtual machines in the clients virtual private cloud. Hard to replicate experiments in different infrastructures.
&
lacks support for the repeatability and reproducibility of distributed experiments. Computing resources vary over time and hence between accesses. Replicability in different infrastructures is not easy to set up.  
\\ \hline
\end{tabular}
\end{table*}

We briefly discuss the limitations of state-of-the-art collaborative environments, with a focus on the specific challenges of the Computing Continuum.

\paragraph{\textbf{Google Colab~\cite{colab}}} 
Mainly used by the AI community (more than 50K users), it is a ready-to-use Jupyter notebook service. Colab notebooks are stored in the \emph{.ipynb} open-source Jupyter notebook format~\cite{colab-faq}, and come with the most popular AI libraries and frameworks installed (\emph{e.g.,} Scikit-Learn~\cite{pedregosa2011scikit}, TensorFlow~\cite{abadi2016tensorflow}, PyTorch~\cite{paszke2019pytorch}, \emph{etc.}) and allow users to run python code through the browser. It is typically used for machine learning, data analysis and education. Colab is popular because it allows users to share Jupyter notebooks without having to download, install, or run anything. Besdides, it provides free access to very expensive computing resources such as GPUs and TPUs. Colab permits multiple users to collaborate on the same notebook. Sharing datasets, ML models, pipelines, and notebooks on AI Hub~\cite{aihub} is also possible (more than 167 notebooks). Its GitHub integration allows users to quickly open GitHub-hosted Jupyter notebooks in Google Colab.

\paragraph{\textbf{Kaggle~\cite{kaggle}}} 
This is a data science and AI platform that offers a customizable Jupyter notebook environment. Kaggle is a subsidiary of Google and, like Colab, it provides free access to GPUs as well as a repository of community-published (more than 10.3 million users) datasets (more than 50K public datasets) and code (\emph{e.g.,} machine learning code) with more than 400K public notebooks. Kaggle is integrated with AI Hub and is popular in the data science and machine learning communities. Kaggle is also well-known for promoting Community Competitions in machine learning at no cost. The main differences~\cite{kaggle-vs-colab} between Colab and Kaggle are: (1) Kaggle allows collaboration with other users on its Web site, while Colab allows collaboration with anyone using the notebook link; (2) Kaggle has a lot of data sets that users can use directly (\emph{e.g.,} notebooks already set up with Kaggle databases~\cite{kaggle-datasets}), while in Colab setting up notebooks with Google Drive~\cite{colab-drive} or managing files~\cite{gsutil-tool} (\emph{e.g.,} to load data sets, files, and images) requires extra work; and (3) Kaggle creates a history of notebook commits that we can  be reviewed.

\paragraph{\textbf{Code Ocean~\cite{clyburne2019computational}}} 
Designed according to FAIR~\cite{wilkinson2016fair} (\emph{i.e.,} Findable, Accessible, Interoperable, and Reusable), Code Ocean aims to make scientific work reproducible. It introduces the concept of Compute Capsule, which refers to Docker~\cite{docker} containers composed of code, data, environments, and results. Capsules provide ready-to-use tools such as Git, Jupyter, RStudio, among others. Its integration with Git allows users to save changes on capsules and then commit them with just one click. Furthermore, users can easily share the link of a capsule and grant permissions. Code Ocean provides scalable compute and storage resources hosted on Amazon Web Services. Resources used by capsules are scaled out when the demand exceeds the machine capacity. Finally, Code Ocean provides a public Capsule Repository~\cite{code-ocean} with more than 1K research capsules. It allows authors of an article to incorporate capsules into the submission process via a Hub publishing API.

\begin{framed}
\vspace{-0.1cm}

Despite these systems being widely used by the AI and data science communities, they present some limitations that hinder their adoption for Computing Continuum research. Table~\ref{tbl:limitations} summarizes these limitations in terms of: 

\begin{enumerate}
    \item access to heterogeneous computing resources, from the IoT/Edge to the Cloud/HPC;
    \item support for large-scale experimental evaluations;
    \item repeatability and reproducibility of experiments on the same hardware setup, and replicability on different infrastructures.
\end{enumerate}

In summary, collaborative environments lack support for providing access to heterogeneous resources (\emph{e.g.,} Edge-to-Cloud); performing experiments at large-scale; and achieving the repeatability, reproducibility, and the replicability of experiments in different testbeds. Hence, the need for novel approaches for reproducible evaluations of workflows targeting the characteristics of the Computing Continuum. 

\vspace{-0.1cm}
\end{framed}

\section{Kheops Design}
\label{sec:kheops}
\begin{figure}[t]
    \lstset{aboveskip=0pt,belowskip=0pt}
\begin{lstlisting}[language=Python, escapechar=|, caption=E2Clab: layers and services configuration. Hardware details described in Section~\ref{subsec:xp-setup-desc}., label=lis-e2c-layers]
environment:
  g5k:|\label{line:g5k}| cluster: dahu
  iotlab:|\label{line:iotlab}| cluster: grenoble
layers:
- name: cloud
  services:
  - name: Server|\label{line:server}|
    environment: g5k, quantity: 1
- name: edge
  services:
  - name: Client|\label{line:client}|
    environment: iotlab, archi: rpi3, quantity: 5
\end{lstlisting}
\end{figure}

This section introduces KheOps, a collaborative environment for the cost-effective reproducibility and replicability of Edge-to-Cloud experiments. KheOps is designed to meet the experimental requirements of both authors and readers as presented in Section~\ref{sec:introduction}.

\subsection{Architecture and implementation}
\label{subsec:kheops_arch_impl}

\begin{figure*}[t]
    \centering
         \includegraphics[width=0.85\linewidth]{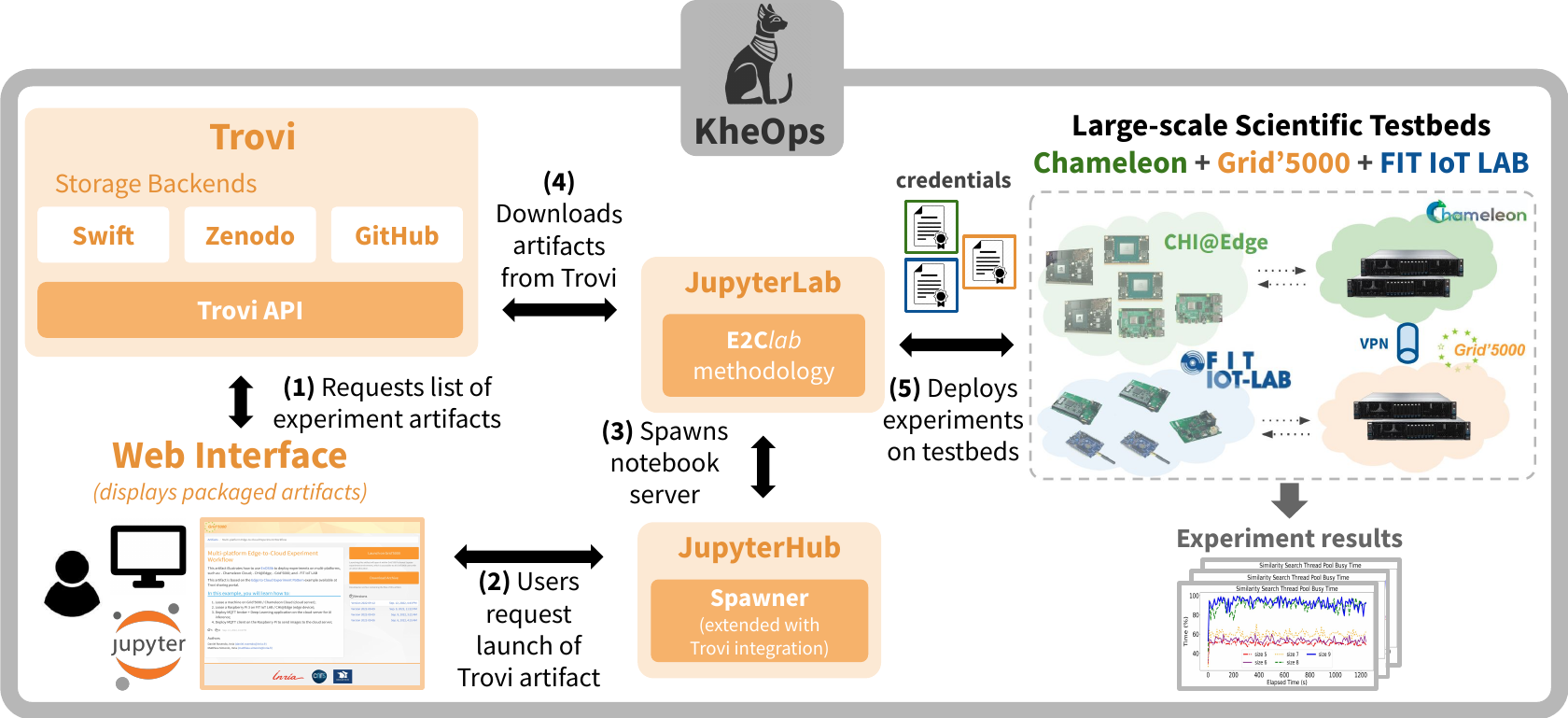}
    \caption{KheOps architecture and experimental workflow.}
    \label{fig:kheops-archi}
\end{figure*}

Figure~\ref{fig:kheops-archi} presents the architecture of KheOps, which consists of three main components: \emph{(i)} Trovi sharing portal; \emph{(ii)} Jupyter environment (JupyterHub service and JupyterLab server); and \emph{(iii)} E2Clab framework (multi-platform experiment methodology). Next, present the integration details of KheOps three components, and we briefly describe their main roles.

\begin{figure}[t]
    \lstset{aboveskip=0pt,belowskip=0pt}
\begin{lstlisting}[language=Python, escapechar=|, caption=E2Clab: user-defined service for the Cloud server., label=lis-e2c-svc]
from e2clab.services import Service
import enoslib as en

class Server(Service):
   def deploy(self):
      with en.actions(roles=self.roles) as a:
        a.shell("pip3 install torchvision torch")
        a.shell("pip3 install pillow paho-mqtt")
      return self.register_service(port=[1883])
\end{lstlisting}
\end{figure}

\subsubsection{Experiment repository} KheOps uses Trovi to share research artifacts such as packaged experiments. These artifacts may be publicly available to allow others to recreate and rerun experiments. Trovi provides a REST API to manage experiment artifacts and integrate them with other systems. The JupyterHub in KheOps uses the Trovi REST API to download artifacts and launch them in the JupyterLab server. 

Artifacts hosted in Trovi can also provide references to repositories like container registries (\emph{e.g.,} DockerHub~\cite{dockerhub}), multipurpose repositories (\emph{e.g.,} Zenodo~\cite{zenodo}), code repositories (\emph{e.g.,} Github~\cite{github}), and among others.

\subsubsection{Notebook environment} Following our previous work~\cite{anderson2019case} on integrating experiment workflows with Jupyter notebooks, we extend JupyterHub to authenticate users and to download (using the Trovi REST API) the experiment artifacts available at Trovi. We also extend JupyterLab to allow users to easily share their experiments in Trovi. Furthermore, JupyterLab is set up with the E2Clab framework as an experimental methodology.

The JupyterLab is packaged with code, data, environment configurations, and experiment results. Its notebooks (file extension \textit{.ipynb}) allow users to run experiments step-by-step by combining text (\emph{e.g.,} explaining the reasoning of the experiments: \emph{What} parameters? \emph{Why} these parameters? and \emph{How} it was set up?) with executable code. Such notebooks are ready to use (\emph{e.g.,} installed with required library/software), executed through a browser, and shared as a Trovi artifact.

\subsubsection{Multi-testbed experiment methodology} KheOps uses the E2Clab methodology to deploy experiments on large-scale scientific testbeds such as Grid'5000, Chameleon Cloud, CHI@Edge, and FIT IoT LAB. Notebooks come with three main template files (\emph{e.g.,} executable code cells in the notebook, presented in Listings~\ref{lis-e2c-layers} to~\ref{lis-e2c-workflow}) that users can benefit from to easily configure and adapt the deployment logic (\emph{e.g.,} computing resources, network, and application execution) according to their experimental needs. 

The first file, named \emph{layers\_services.yaml} and presented in Listing~\ref{lis-e2c-layers}, allows users to lease IoT/Edge and Cloud/HPC resources. Through this file, users may also set up their applications and services as presented in Listing~\ref{lis-e2c-svc}. Next, the \emph{network.yaml} file (Listing~\ref{lis-e2c-net}) allows users to define delay, loss, and bandwidth between computing resources. Finally, the \emph{workflow.yaml} file (Listing~\ref{lis-e2c-workflow}) guides users to define the experiment workflow through three main steps: \emph{prepare} (\emph{e.g.,} copy artifacts to remote nodes, install libraries, \emph{etc.}), \emph{launch} (\emph{e.g.,} execute the application parts), and \emph{finalize} (\emph{e.g.,} backup results from remote nodes to the JupyterLab server).

E2Clab abstracts all the complexities of deploying and executing experiments across various testbeds. To do so, users need to add the credential files of the respective testbeds to their notebooks. Setting up a VPN is also supported as this may be required to enable the communication between different geographically distributed tesbeds (\emph{e.g.,} Chameleon in the USA and Grid'5000 in France).

\begin{figure}[t]
    \lstset{aboveskip=0pt,belowskip=0pt}
\begin{lstlisting}[language=Python, escapechar=|, caption=E2Clab: network configuration., label=lis-e2c-net]
networks:
- src: cloud, dst: edge
  delay: "150ms", rate: "25kbit", loss: "0.02"
\end{lstlisting}
\end{figure}

\subsection{Experimental workflow}
\label{subsec:kheops_workflow}

In summary, the workflow for launching an experiment artifact on large-scale testbeds consists of 5 main steps. First, through a web interface, users can browse the list of experimental artifacts publicly available in Trovi (step 1). Selecting an artifact displays details such as the experiment description, the authors and contact information, and the artifact versions. 

A \emph{launch} button allows users to execute the artifact (step 2). This button redirects users to the JupyterHub service. After authentication, the request to launch the artifact is sent to the JupyterHub Spawner. Next, the Spawner spawns the JupyterLab server (step 3)  and then it downloads experimental artifacts such as notebooks, code, and datasets, among others (step 4). The JupyterLab service is set up with the E2Clab framework as the experimental methodology. Finally, users can execute the code cells from the notebook to lease IoT/Edge and Cloud/HPC computing resources available on the testbeds, deploy and execute the application, and gather the experiment results (step 5).

\begin{figure}[t]
    \lstset{aboveskip=0pt,belowskip=0pt}
\begin{lstlisting}[language=Python, escapechar=|, caption=E2Clab: workflow configuration., label=lis-e2c-workflow]
- hosts: edge.*
  depends_on:
    conf_selector: cloud.server.*
    grouping: round_robin, prefix: "server"
  prepare:
  - copy:
      src:{{working_dir}}/artifacts, dest: /
  launch:
  - shell: bash /edge_worker.sh edge_data 100 "{{server.ip}}" False
  finalize:
  - fetch:
      src:/tmp/predict.log, dest:{{working_dir}}/
\end{lstlisting}
\end{figure}

Steps 2 to 4 are automatically executed. This is a one-click feature that allows users to have a ready-to-use environment for reproducing and replicating complex Edge-to-Cloud experiments in a cost-effective manner. Note that the whole workflow requires only three clicks: selecting the experiment artifact (step 1); then launching it (steps 2 to 4); and executing it on the testbeds (step 5).

\section{Evaluation}
\label{sec:evaluation}

\begin{figure}[t]
    \centering
         \includegraphics[width=0.93\linewidth]{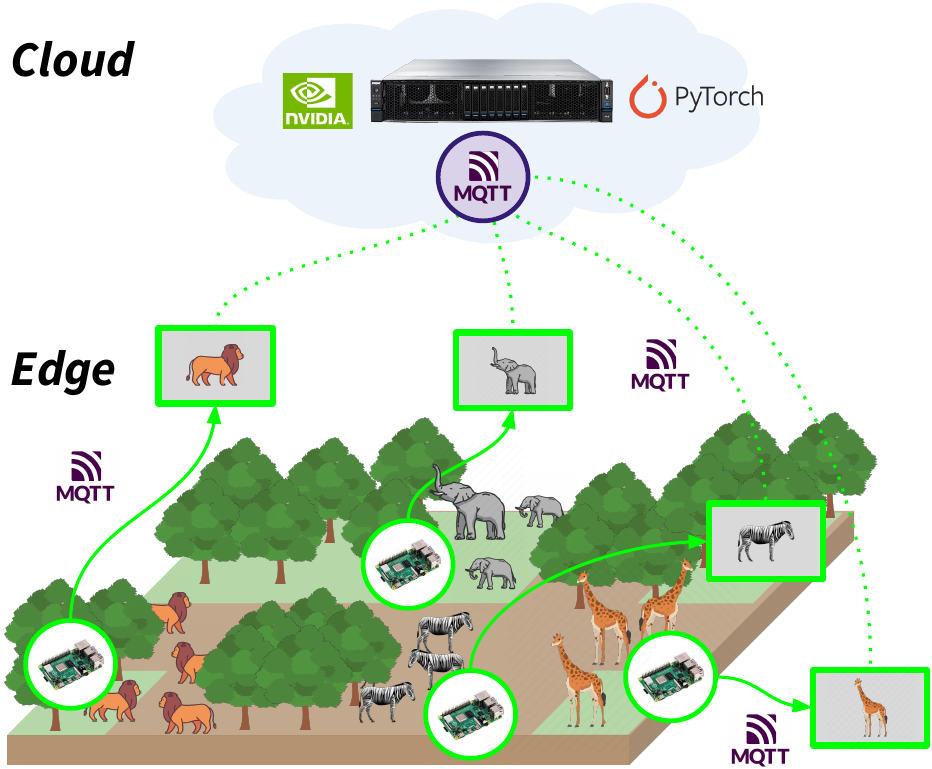}
    \caption{Edge-to-Cloud application: monitoring animals migration in the African savanna.}
    \label{fig:xpsetup}
\end{figure}

In this section, we show how KheOps can be used to analyze the performance of a real-life Edge-to-Cloud application deployed in the African savanna (illustrated in Figure~\ref{fig:xpsetup}). This application is composed of distributed Edge devices monitoring animal migration in the Serengeti region. Devices at the Edge collect and compress wildlife images, then the image is sent to the Cloud where the animal classification happens using a pre-trained Neural Network model. Finally, classified data helps conservationists to learn what management strategies work best to protect species.

The goals of these experiments are:

\begin{itemize}
    \item to understand the impact on performance of \textbf{Cloud-centric} and \textbf{Hybrid (Edge+Cloud)} processing;
    \item to show how authors of an article can benefit from KheOps to make their \textbf{experiments reproducible};
    \item to show how readers of an article can leverage KheOps to \textbf{replicate the experiments} in an article (published using KheOps). 
\end{itemize}

\begin{figure*}[t]

\begin{subfigure}{.25\textwidth}
  \centering
  \includegraphics[width=\linewidth]{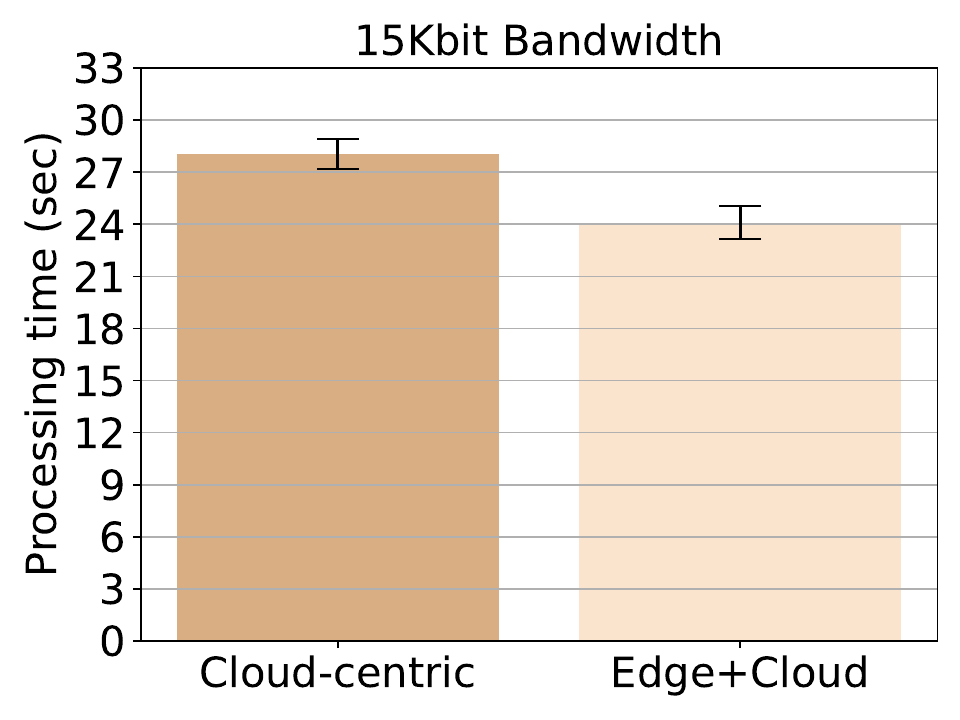}
  \caption{Grid'5000 + FIT IoT LAB.}
  \label{fig:15g5k}
\end{subfigure}%
\begin{subfigure}{.25\textwidth}
  \centering
  \includegraphics[width=\linewidth]{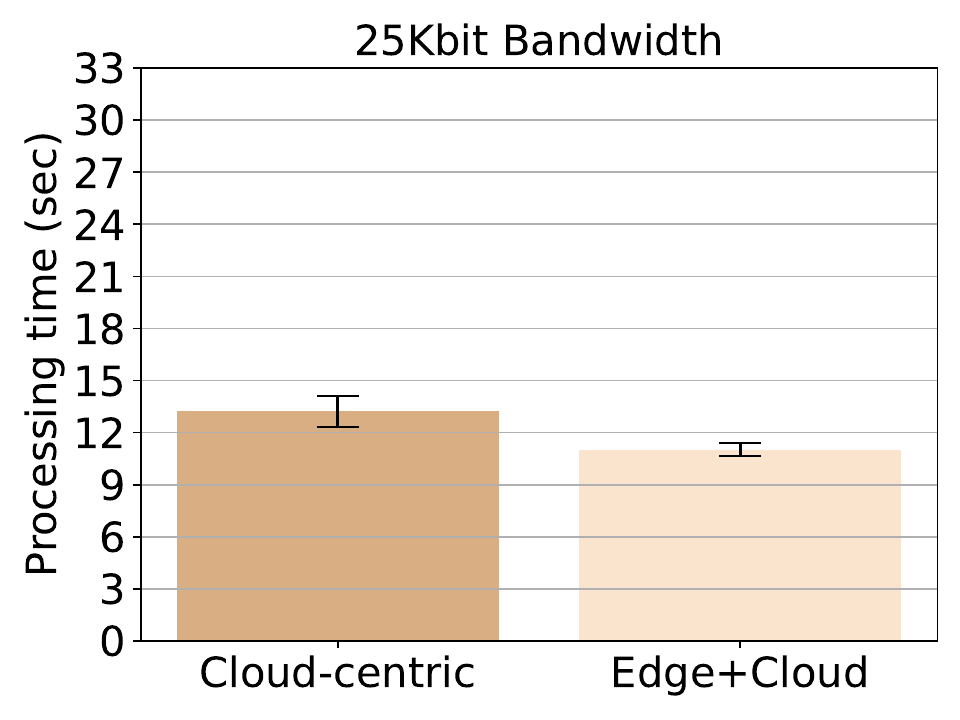}
  \caption{Grid'5000 + FIT IoT LAB.}
  \label{fig:25g5k}
\end{subfigure}%
\begin{subfigure}{.25\textwidth}
  \centering
  \includegraphics[width=\linewidth]{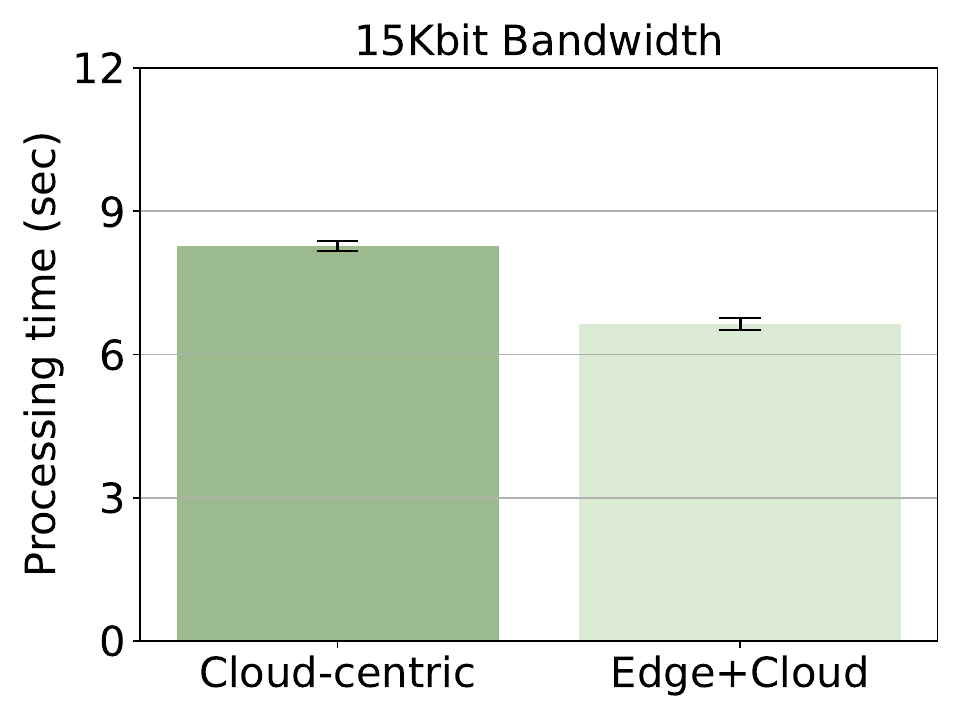}
  \caption{Chameleon + CHI@Edge.}
  \label{fig:15chi}
\end{subfigure}%
\begin{subfigure}{.25\textwidth}
  \centering
  \includegraphics[width=\linewidth]{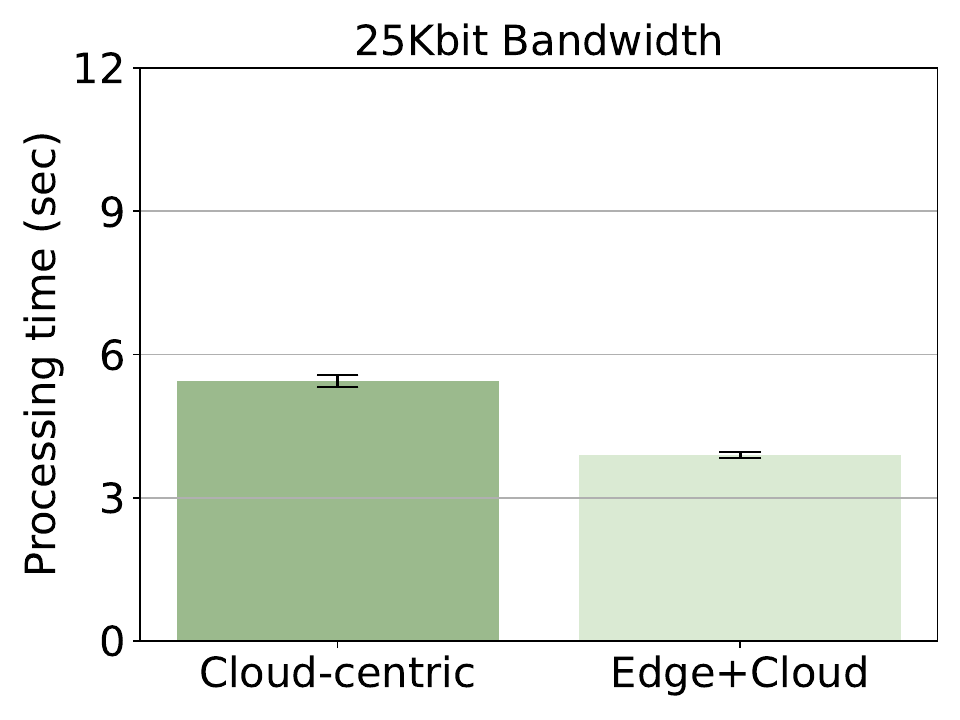}
  \caption{Chameleon + CHI@Edge.}
  \label{fig:25chi}
\end{subfigure}%
\caption{Cloud-centric vs Edge+Cloud processing: (a, b) executed by authors on Grid'5000 and FIT IoT LAB testbeds; and (c, d) replicated by readers on Chameleon Cloud and CHI@Edge testbeds.}
\label{fig:overhead_analysis}
\end{figure*}

\begin{figure}[t]
\begin{subfigure}{.25\textwidth}
  \centering
  \includegraphics[width=\linewidth]{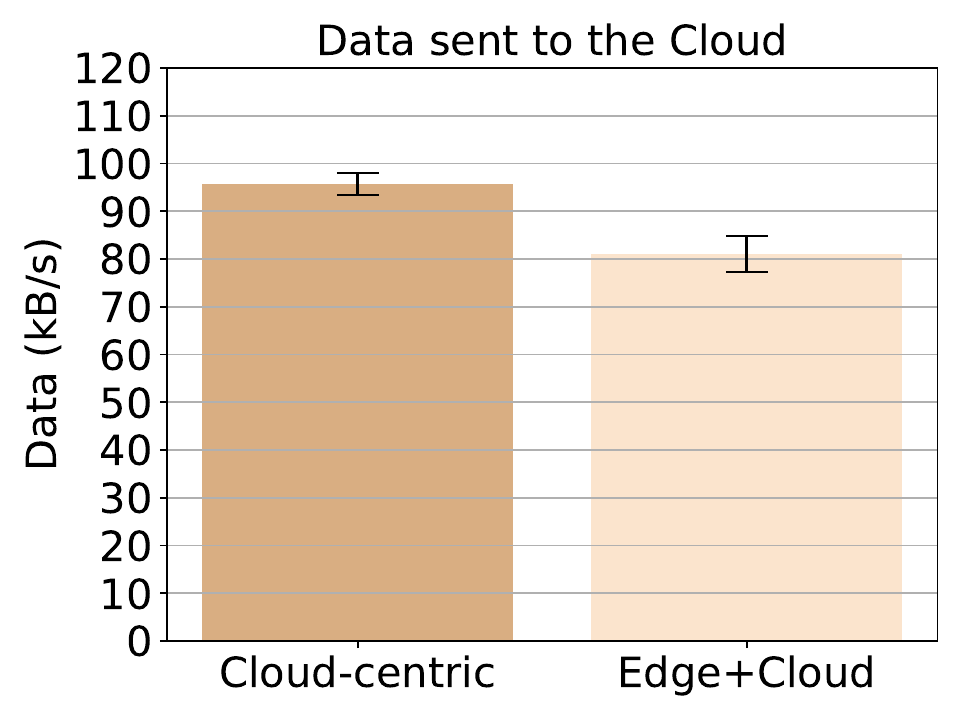}
  \caption{Grid'5000 + FIT IoT LAB.}
  \label{fig:net-g5k}
\end{subfigure}%
\begin{subfigure}{.25\textwidth}
  \centering
  \includegraphics[width=\linewidth]{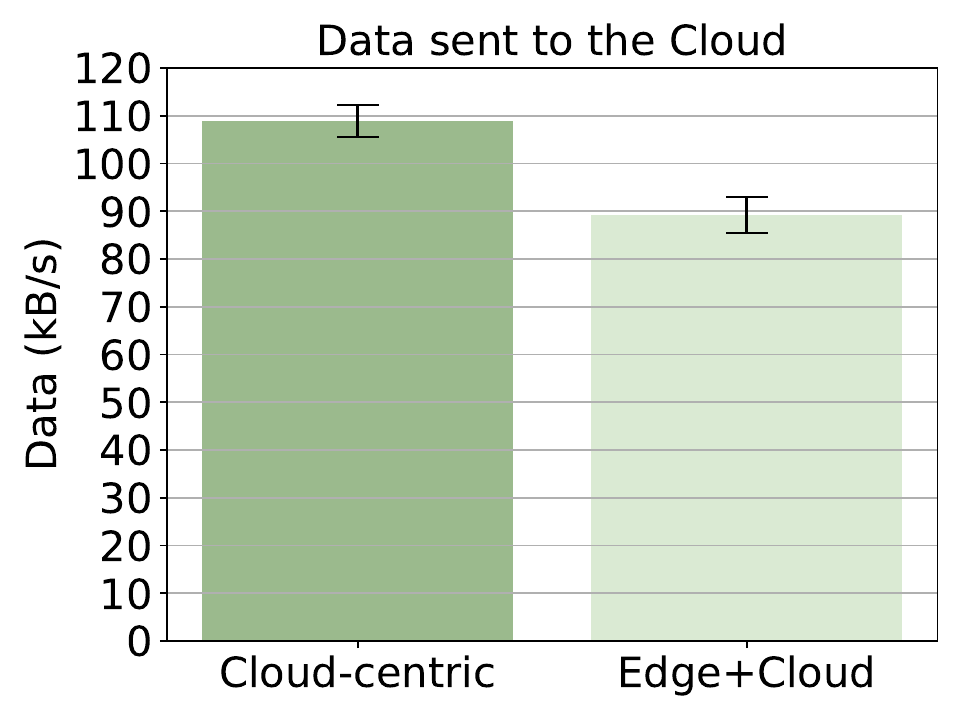}
  \caption{Chameleon + CHI@Edge.}
  \label{fig:net-chi}
\end{subfigure}%
\caption{Amount of data sent to the Cloud regarding the Cloud-centric and Edge+Cloud processing approaches.}
\label{fig:monitoring_net}
\end{figure}


\begin{figure*}[t]

\begin{subfigure}{.25\textwidth}
  \centering
  \includegraphics[width=\linewidth]{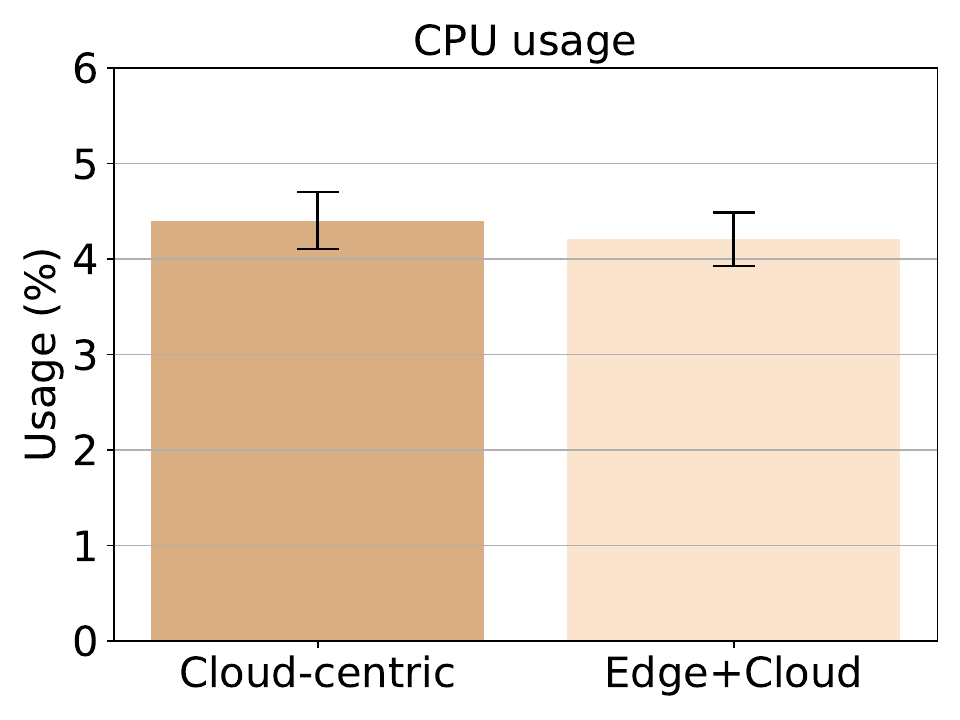}
  \caption{Grid'5000 + FIT IoT LAB.}
  \label{fig:cpu-g5k}
\end{subfigure}%
\begin{subfigure}{.25\textwidth}
  \centering
  \includegraphics[width=\linewidth]{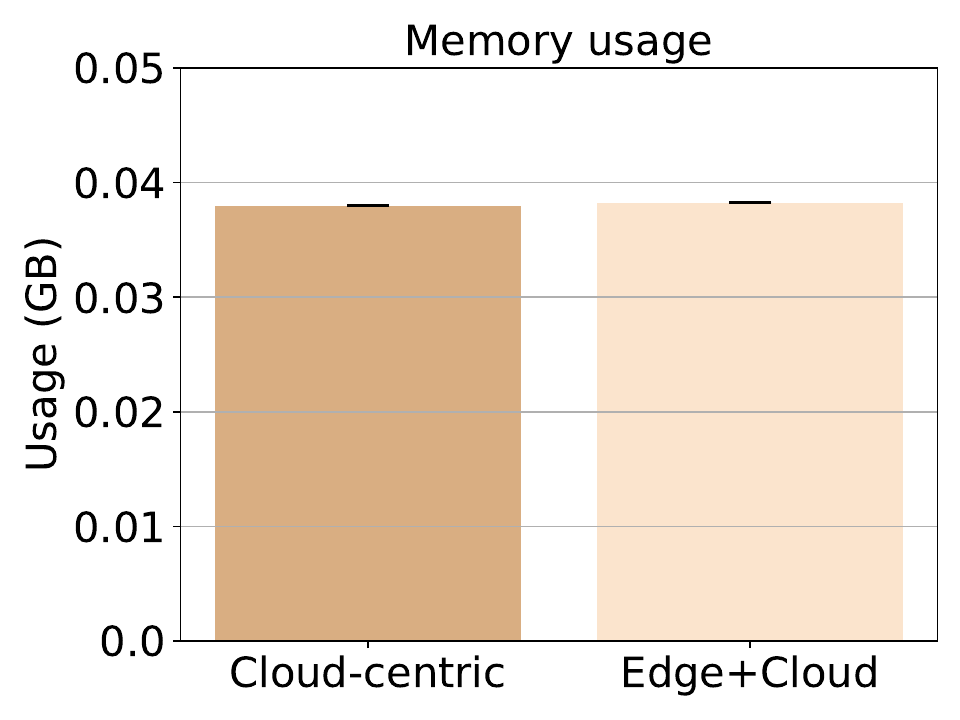}
  \caption{Grid'5000 + FIT IoT LAB.}
  \label{fig:mem-g5k}
\end{subfigure}%
\begin{subfigure}{.25\textwidth}
  \centering
  \includegraphics[width=\linewidth]{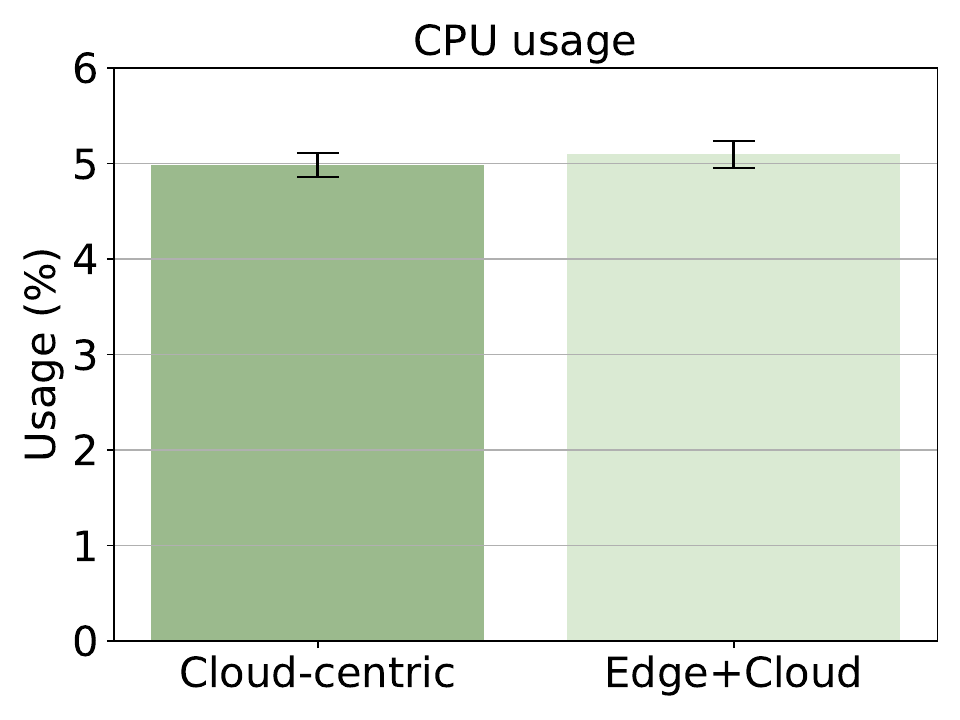}
  \caption{Chameleon + CHI@Edge.}
  \label{fig:cpu-chi}
\end{subfigure}%
\begin{subfigure}{.25\textwidth}
  \centering
  \includegraphics[width=\linewidth]{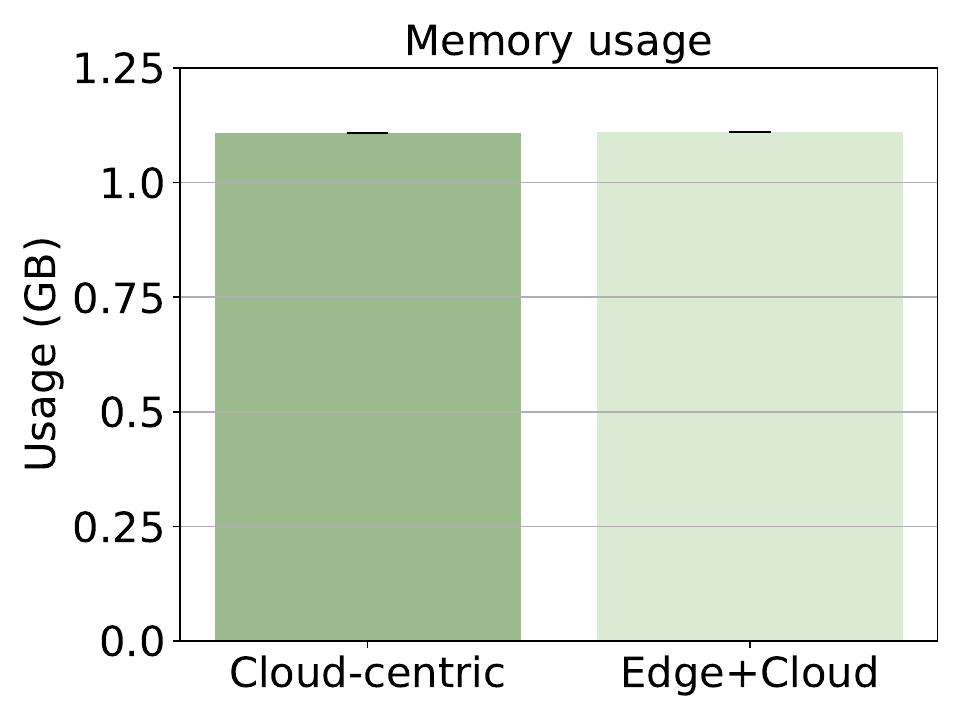}
  \caption{Chameleon + CHI@Edge.}
  \label{fig:mem-chi}
\end{subfigure}%
\caption{Resource consumption on the Edge device: CPU and Memory usage.}
\label{fig:monitoring_rec}
\end{figure*}



To reproduce the evaluations in this section, refer to~\cite{xp-artifacts-trovi}.

\subsection{Experimental setup}
\label{subsec:xp-setup-desc}

\paragraph{\textbf{Application performance metrics}} 
The main tracked metric is the \emph{processing time}, which refers to the time required to: pre-process the image captured (\emph{e.g.,} image compression on the Edge device); transmit the image to the Cloud server; and finally decompress the image and predict the animal through an AI model. In addition, we analyze the \emph{amount of data transmitted} to the Cloud and the \emph{resource consumption} (\emph{e.g.,} CPU and memory) on the Edge device.

To increase the accuracy of the results, we measure the processing duration 100 times for each experiment, each time with a different image and an interval of 30 seconds (\emph{i.e.,} Edge devices transmit images to the Cloud server every 30 seconds). The remaining metrics are captured using Dool (Dstat)~\cite{dool} at application runtime. All results are presented as the mean followed by their respective 95\% confidence interval.

\paragraph{\textbf{KheOps replicability metric}} 
To measure how close/precise readers experiments are from authors experiments, we define the Replicability Accuracy ($Rep_{accuracy}$) metric. For assessing variability and error in results~\cite{national2019reproducibility}, a recommendation is to repeat the experiments multiple times to achieve narrower inferential error bars (i.e., confidence interval, standard deviation, \emph{etc.})~\cite{cumming2007error}. The Replicability Accuracy metric is calculated as Equation~\ref{eq:rep_acc}:

\begin{equation}
\label{eq:rep_acc}
\begin{aligned}
Rep_{accuracy} = 1 - \bigg | 
\frac{min(x_{1A}, x_{2A})}{max(x_{1A}, x_{2A})} - 
\frac{min(x_{1R}, x_{2R})}{max(x_{1R}, x_{2R})} \bigg | 
\end{aligned}
\end{equation}

Ideally, $Rep_{accuracy}$ would be close to $1$. $x_{iA}$ and $x_{iR}$ refer to the application performance metric value obtained from authors and readers experiments, respectively. For instance, in Figure~\ref{fig:15g5k}, $x_{1A}$ refers to the Cloud-centric bar and $x_{2A}$ to the Edge+Cloud bar.


\paragraph{\textbf{Workload}}
Devices at the Edge transmit images (from the Snapshot Serengeti dataset~\cite{usecase-dataset} composed of millions of wildlife images collected annually) to the Cloud server that predicts animals using a trained MobileNetV3 Convolutional Neural Network model. We evaluate this workload considering two network configurations, 25Kbit and 15Kbit bandwidth with a round-trip delay of 150ms.

\paragraph{\textbf{Software}} On the Edge devices, we use the zlib~\cite{zlib} Python library to compress images. MQTT~\cite{mqtt} protocol is used to transmit images to the Cloud server. On the Cloud server, we use an MQTT broker to receive images, then zlib to decompress images, and finally PyTorch to predict animals.

\paragraph{\textbf{Hardware}} 
The \textbf{authors} perform experiments on the following testbeds in France: Grid’5000 and FIT IoT LAB. On Grid'5000 (Cloud server), they use the dahu~\cite{g5kdahu} machine equipped with an Intel Xeon Gold 6130 CPU 2.10GHz, 16 cores/CPU, 192GB of RAM, and Ethernet network. On FIT IoT LAB (Edge device), they use a Raspberry Pi 3 Model B~\cite{fitrpi3} with four ARM Cortex-A53 processing cores running at 1.2GHz, 1GB LPDDR2 memory, and 2.4GHz 802.11ac wireless LAN. 

The \textbf{readers} replicate authors experiments on the following testbeds in USA: Chameleon Cloud and CHI@Edge. On Chameleon CHI@TACC (Cloud server), they use the Skylake~\cite{chitacc} machine equipped with an Intel Xeon Gold 6126 CPU 2.60GHz, 12 cores/CPU, 192GB of RAM, and Ethernet network. On CHI@Edge (Edge device), they use a Raspberry Pi 4~\cite{chiedgerpi4}, with four BCM2711 Cortex-A72 processing cores running at 1.5GHz, 8GB LPDDR4 memory, and 2.4GHz and 5GHz 802.11ac wireless LAN.

\subsection{\textbf{How KheOps helps experiment authors}} 

Let us consider the requirements of the experiment authors (a-REQ) as introduced in Section~\ref{sec:introduction}.

\paragraph{\textbf{a-REQ 1. Execute experiments on heterogeneous computing resources}} KheOps provides access to IoT/Edge devices and Cloud/HPC resources at large-scale, using the E2Clab methodology. Supported testbeds include (but are not limited to, as explained in Section~\ref{subsec:integration}): Grid'5000, FIT IoT LAB.

\paragraph{\textbf{a-REQ 2. Systematically describe and explain the experimental processes and their reasoning}} Through Jupyter notebooks and the E2Clab configuration files, the authors describe and explain the experiment design choices such as the layers (\emph{e.g.,} Edge and Cloud), the services (\emph{e.g.,} the Edge client and the Cloud server), the network constraints, and the application workflow execution. This is done in Jupyter notebooks by combining text (explaining the configurations) followed by executable code (E2Clab files).

\paragraph{\textbf{a-REQ 3. Efficiently configure the experimental infrastructure and repeat the experiments}} All the complexities of configuring the Edge-to-Cloud infrastructure, such as leasing computing resources, mapping the application parts (\emph{e.g.,} Edge and Cloud services), enforcing the network constraints, and executing the workflow are transparently handled by KheOps. The authors just need to define their experimental needs in the E2Clab configuration files. Repeating and adapting the experiments (\emph{e.g.,} changing the network constraints) is easily done through E2Clab instrumentation. 

\paragraph{\textbf{a-REQ 4. Easily share the experiment artifacts in a public and safe repository}} Through the Trovi and JupyterLab integration, authors can upload their artifacts to the Trovi sharing portal with a few clicks.\\

We discuss the experimental results from the authors perspective, using the three application performance metrics mentioned earlier. 

\subsubsection{\textbf{Impact of the network on the processing time}} 
Authors define two sets of experiments. In the first one (Figure~\ref{fig:15g5k}), they fix the network bandwidth at 15Kbit and vary the processing approach between Cloud-centric and Hybrid (Edge+Cloud). In the second one (Figure~\ref{fig:25g5k}), they fix the bandwidth at 25Kbit for both processing approaches.

From the results, the authors observe that the Hybrid (Edge+Cloud) approach outperforms the Cloud-centric one for both network configurations. In the 15Kbit bandwidth setup, the processing time for the Cloud-centric is about 27 seconds on average, against 24 seconds for the hybrid processing. In the 25Kbit bandwidth configuration, this difference is lower, 13 seconds and 11 seconds for the Cloud-centric and Hybrid, respectively. The higher the bandwidth, the lower will be the difference between the two processing approaches. This is because image transmission is the most time-consuming task among the other tasks (\emph{i.e.,} compressing/decompressing images and model inference).

\subsubsection{\textbf{Amount of data sent to the Cloud}} According to the results presented in Figure~\ref{fig:net-g5k}, authors observe that the Hybrid (Edge+Cloud) approach transmits less data (81kB/s on average) to the Cloud compared to the Cloud-centric approach (96kB/s on average). This is because, in Hybrid processing, Edge devices compress images before transmitting them to the Cloud.

\subsubsection{\textbf{Resource consumption on the Edge device}} 
Results in Figures~\ref{fig:cpu-g5k} and~\ref{fig:mem-g5k} show that there is no significant difference in the CPU and memory usage in the Edge device when changing between the Cloud-centric and Hybrid processing approaches. CPU usage is around 4.2\% and 4.4\% for Hybrid and Cloud-centric processing, respectively. Memory usage is around 0.38GB for both.

\subsection{\textbf{How KheOps helps readers}} 
After the authors publish their results, other researchers from a different lab download the article from a scientific database and decide to replicate the study on their own premises (\emph{e.g.,} on a different testbed). Following the same logic, we present how KheOps helps the readers to replicate the experiments cost-effectively, that is, according to the readers requirements (r-REQ) in Section~\ref{sec:introduction}.

\paragraph{\textbf{r-REQ 1. Find and access the experiment as simply as finding and reading the paper}} 
Through the KheOps web interface (step 1 in Figure~\ref{fig:kheops-archi}) the readers obtain access to all the public experiments shared by the community and available in Trovi. Then, they select the experiment shared by the authors of the article to get more details.

\paragraph{\textbf{r-REQ 2. Perform the experiment, not just read about it}} 
Next, in the experiment details web page,  readers can launch a JupyterLab server with artifacts in just a single click (steps 2, 3, and 4 in Figure~\ref{fig:kheops-archi}). Finally, following the experiment instructions described in the Jupyter notebook, the readers deploy and execute the experiments on their testbeds, such as (but not limited to): the Chameleon Cloud and CHI@Edge (step 5 in Figure~\ref{fig:kheops-archi}). 

\paragraph{\textbf{r-REQ 3. Experiment reasoning: \emph{“What”}, \emph{“Why”}, and \emph{“How”}}} Before running the experiments, the readers can go through the Jupyter notebook to understand \emph{What} the experiment does (\emph{e.g.,} capture and compress images on Edge devices and then decompress the images and predict the animals on the Cloud server). The readers can also discover \emph{Why} the authors set up the experiment with a 25kbit and 15kbit network bandwidth. Finally, KheOps allows to understand \emph{How} the authors interconnect the Edge devices with the Cloud server (\emph{e.g.,} assigning a public IP to the Cloud server, or opening firewall rules; using the MQTT protocol; among others).

\paragraph{\textbf{r-REQ 4. Efficiently configure the experimental infrastructure}} 
To achieve this, the readers just have to adapt the \emph{layers\_services} configuration file (presented in Listing~\ref{lis-e2c-layers}) to the Chameleon Cloud and CHI@Edge testbeds. Configuring the network bandwidth to 25kbit and then changing it to 15kbit is as simple as changing the \emph{rate} parameter in the \emph{network} file (Listing~\ref{lis-e2c-net}). Finally, copying data to the Edge device, interconnecting it with the Cloud server, launching the application, and finally collecting the results is as simple as defining the \emph{workflow} configuration file (Listing~\ref{lis-e2c-workflow}). The \emph{network} and \emph{workflow} configuration files are testbed agnostic, meaning that users do not need to update these files when changing the deployment from Grid'5000 + FIT IoT LAB to Chameleon + CHI@Edge.\\

Next, we report on the replicated experiments. 

\begin{table}[t]
\small
\centering
\caption{Accuracy of replicated experiments.}
\label{tbl:accuracy}
\begin{tabular}{lll}
\hline
\textbf{Metric}        & \textbf{\begin{tabular}[c]{@{}l@{}}Replicability \\ accuracy\end{tabular}} & \textbf{Experiment result}                                                                                                         \\ \hline
\rowcolor[HTML]{D9D9D9} 
Processing time 15Kbit        & 0.943                                                                      & \cellcolor[HTML]{D9D9D9}Figure~\ref{fig:15g5k} and~\ref{fig:15chi}     \\
Processing time 25Kbit  & 0.882                                                                      & Figure~\ref{fig:25g5k} and~\ref{fig:25chi}                             \\
\rowcolor[HTML]{D9D9D9} 
Data sent to the cloud & 0.973                                                                      & \cellcolor[HTML]{D9D9D9}Figure~\ref{fig:net-g5k} and~\ref{fig:net-chi} \\
CPU usage              & 0.978                                                                      & Figure~\ref{fig:cpu-g5k} and~\ref{fig:cpu-chi}                         \\
\rowcolor[HTML]{D9D9D9} 
Memory usage           & 0.996                                                                      & \cellcolor[HTML]{D9D9D9}Figure~\ref{fig:mem-g5k} and~\ref{fig:mem-chi} \\ \hline
\end{tabular}
\end{table}

\subsubsection{\textbf{Impact of the network on the processing time}} 
From the results in Figures~\ref{fig:15chi} and~\ref{fig:25chi}, readers conclude that the Hybrid (Edge+Cloud) processing approach outperforms the Cloud-centric one for both network configurations. This conclusion is consistent with the results observed in the published article.  

Following the analysis, readers observe that in the 15Kbit bandwidth network configuration, the processing time for the Cloud-centric is about 8 seconds on average, against 6.5 seconds for the hybrid processing. In the 25Kbit bandwidth  setup, this difference is lower, 5.5 seconds and 4 seconds for the Cloud-centric and Hybrid, respectively. Similarly to the authors results, readers also observe that the higher the bandwidth, the lower will be the difference between the two processing approaches.

Furthermore, as presented in Table~\ref{tbl:accuracy}, we highlight that readers obtained a replicability accuracy of 88.2\% and 94.3\% for 15Kbit and 25Kbit network configurations, respectively.

\subsubsection{\textbf{Amount of data sent to the Cloud}} According to the results presented in Figure~\ref{fig:net-chi}, readers observe that the Hybrid approach transmits less data than the Cloud-centric. The former transmits around 89.2kB/s and the latter 108.8kB/s. Compressing images on the Edge helps to reduce the amount of data sent to the Cloud server. This conclusion is also consistent with the published article and presents a replicability accuracy of 97.3\%.

\subsubsection{\textbf{Resource consumption on the Edge device}} 
Results in Figures~\ref{fig:cpu-chi} and~\ref{fig:mem-chi} show that there is no significant difference in the CPU and memory usage between the Cloud-centric and the Hybrid processing approaches. CPU usage is around 5.1\% and 5\% for Hybrid and Cloud-centric processing, respectively. Memory usage is around 1.1GB for both. We highlight that these conclusions are consistent with the published article and present a replicability accuracy of 97.8\% and 99.6\% for CPU and memory usage, respectively.

\begin{framed}
\vspace{-0.1cm}

Despite readers observing a lower processing time compared to the authors, they could verify that their experiment conclusions are consistent with the original study, and their results present a high replicability accuracy (see Table~\ref{tbl:accuracy}). 

This time difference is expected since readers used a more powerful Edge device (Raspberry Pi4 against Raspberry Pi3) for processing the most time-consuming task (\emph{e.g.,} image compression and then transmission). The Raspberry Pi4 has more RAM memory (8GB \emph{vs.} 1GB in Raspberry Pi3), a better CPU (1.5GHz \emph{vs.} 1.2GHz), network (5GHz \emph{vs.} 2.4GHz). 

Furthermore, regarding the remaining metrics such as the amount of data sent to the Cloud and the CPU and memory usage on the Edge device, readers observe small differences when replicating the original study in different testbeds. This is due to the different deployment approaches used by each testbed, for instance, in FIT IoT LAB the Raspberry Pi 3 board runs an embedded Linux that is built with Yocto~\cite{yocto}, while CHI@Edge is based on Docker~\cite{docker} containers. Despite that, the conclusions observed by authors and readers are the same and present high accuracies.

\vspace{-0.1cm}
\end{framed}


\section{Discussion}
\label{sec:discussion}



KheOps core elements (\emph{i.e.,} Trovi, JupyterLab, E2Clab)  exhibit several features that make it a promising environment for advancing Computing Continuum research through reproducible and replicable experiments. We briefly discuss them here.

\subsection{Usability and reusability} 

KheOps targets \textbf{usability} by allowing users to easily find experiment artifacts shared in Trovi and then to launch experiments in a JupyterLab server in just a few clicks. KheOps abstracts all the low-level details of defining and configuring the experimental environment. It provides a high-level abstraction for mapping application parts with the Edge and Cloud infrastructures. Besides, the configuration files used to define the whole experimental environment are designed to be easy to use and understand. 

KheOps also targets \textbf{reusability} of the experiment artifacts. For instance, readers of an article can reuse the authors artifacts to replicate the study or build upon the existing artifacts to generate new results. In addition, through E2Clab \emph{User-Defined Services}, users can define their own services (\emph{e.g.,} the Edge client and the Cloud server) with the desired deployment logic (\emph{e.g.,} mapping the services to the physical machines/devices; installing required software and packages; \emph{etc.}). Such services can be shared in this repository~\cite{e2clab-uds}. 

\subsection{Analyzing other real-life applications}

The KheOps approach is \textbf{generic} in terms of deployment and analysis of \textbf{other applications}. We highlight that, despite our evaluations focusing on the African savanna use-case, KheOps can be easily used in other contexts. Supporting new applications can be achieved by describing and implementing their logic in the \emph{User-Defined Services} configuration file.

\subsection{Integration with other scientific testbeds}
\label{subsec:integration}

The KheOps approach is \textbf{generic} with respect to the \textbf{deployment testbeds}. KheOps allows users to analyze application workflows on various large-scale scientific testbeds, beyond the four testbeds used in this work. The definition of the experimental environment through E2Clab configuration files (\emph{e.g.,} \emph{layer\_services.yaml, network.yaml}, and \emph{workflow.yaml}) is tesbed agnostic, meaning that a deployment on the Grid'5000 testbed can be easily replicated in Chameleon (if the required computing resources are available).

\subsection{Reproducibility and artifact availability}

The experimental evaluations presented in this work follow a rigorous methodology~\cite{rosendo:hal-02916032} to support reproducible Edge-to-Cloud experiments on large-scale scientific testbeds. All the experiment artifacts are publicly available~\cite{xp-artifacts-trovi} at the Trovi sharing portal and the results are also publicly available~\cite{exp-artifacts} in our GitLab repository.

\subsection{KheOps limitations}
Next, we discuss future research under the KheOps approach to help with experiment reproducibility.

\paragraph{Provenance data capture} It may assist in the processes of reproducing complex Edge-to-Cloud workflows~\cite{liu2009encyclopedia}. Typically, users have to execute and repeat various experiments. The output of this process generates hundreds of data related to the experimental setup (\emph{e.g.,} hardware, software, code, data set, \emph{etc.}) and application workflow execution. Analyzing such data is only possible with the help of provenance data capture~\cite{souza_keeping_2019}.

\paragraph{Abstract hardware description} The hardware configuration is a significant barrier to reproducibility~\cite{calasanz2023}, especially in complex Edge-to-Cloud deployments comprising heterogeneous computing resources. The description of resources should be in terms of hardware requirements to execute the experiments (\emph{e.g.,} CPU, GPU, memory, disk, and network). The goal is to abstract the hardware resource description among various testbeds, preventing independent researchers from knowing about the infrastructure of the original experimental environment.

\section{Related Work}
\label{sec:related-work}

We have not found in the literature related work proposing collaborative environments with a focus on the Edge-to-Cloud Continuum. Closer solutions to KheOps, but without focusing on the Computing Continuum, are Google Colab~\cite{colab}, Kaggle~\cite{kaggle}, and Code Ocean~\cite{clyburne2019computational} as presented in Section~\ref{sec:limitations}.

KheOps differs from Google Colab, Kaggle, and Code Ocean mainly regarding the features presented in Table~\ref{tbl:limitations} (limitations) such as the access to heterogeneous Edge-to-Cloud resources; access to large-scale infrastructures; and supporting the experiment repeatability and reproducibility on the same hardware setup and replicability in different infrastructures. In addition, KheOps relies on open scientific testbeds (\emph{e.g.,} Grid5000, FIT IoT LAB, Chameleon, and CHI@Edge) that are highly reconfigurable and controllable and designed to support reproducible experiments.

\section{Conclusion}
\label{sec:conclusions}

KheOps is, to the best of our knowledge, the first collaborative environment supporting the cost-effective reproducibility of applications on the Edge-to-Cloud Continuum. It provides simplified abstractions for systematically defining and explaining the experimental environment through Jupyter notebooks (\emph{e.g.,} infrastructures, services, network, and workflow execution); provides access to heterogeneous computing resources from the IoT/Edge to the Cloud/HPC; and allows researchers to easily find and share the experiment artifacts in the Trovi portal.

The experimental validation shows that KheOps helps authors to make their experiments repeatable and reproducible on the Grid5000 and FIT IoT LAB testbeds. Furthermore, KheOps helps readers to cost-effectively replicate authors experiments in different infrastructures such as Chameleon Cloud + CHI@Edge testbeds, and obtain the same conclusions with accuracies $>$88\% for all performance metrics. 

\section*{Acknowledgments}
This work was funded by Inria through the HPC-BigData Inria Challenge (IPL) and through the UNIFY Associate Team joint in the framework of the JLESC international lab and the HPDaSc associate team with Brazil. It was co-funded by the French ANR OverFlow project (ANR-15- CE25-0003). Experiments presented in this paper were carried out using the Chameleon Cloud, CHI@Edge, Grid'5000, and FIT IoT LAB testbeds, supported by a scientific interest group hosted by several Universities. We also would like to thank Argonne National Laboratory for supporting this work. This material is based upon work supported by the U.S. Department of Energy, Office of Science, under contract number DE-AC02-06CH11357 as well as by the NSF award 2130889 and NIFA award 2021-67021-33775.

\bibliographystyle{ACM-Reference-Format}
\balance
\bibliography{acmart.bib}

\end{document}